\documentclass[%
superscriptaddress,
preprint,
 amsmath,amssymb,
 aps,
]{revtex4-1}

\usepackage{graphicx}
\usepackage{dcolumn}
\usepackage{bm}

\usepackage{mathrsfs}
\usepackage{hyperref}
\hypersetup{
  colorlinks = true,
  urlcolor = blue,
  linkcolor = blue,
  citecolor = green,
  filecolor = magenta,
}

\newcommand{\ud}{\mathrm{d}}
\newcommand{\pd}{\partial}

\newcommand{\tdu}{{\hat{u}}}
\newcommand{\tdr}{{\hat{r}}}
\newcommand{\tdth}{{\hat{\theta}}}
\newcommand{\ta}{{\hat{a}}}
\newcommand{\tb}{{\hat{b}}}
\newcommand{\tc}{{\hat{c}}}

\newcommand{\tdcd}{{\mathscr D}}
\newcommand{\tdgamma}{{\gamma}}

\newcommand{\order}[1]{\mathcal{O}\left(#1\right)}

\newcommand{\tmu}{{\hat{\mu}}}
\newcommand{\tnu}{{\hat{\nu}}}
\newcommand{\trho}{{\hat{\rho}}}
\newcommand{\tsigma}{{\hat{\sigma}}}

\newcommand{\bmu}{{\bar{\mu}}}
\newcommand{\bnu}{{\bar{\nu}}}
\newcommand{\brho}{{\bar{\rho}}}
\newcommand{\bsigma}{{\bar{\sigma}}}

\begin{document}


\title{The general property of the tensor gravitational memory effect in theories of gravity}
\author{Shaoqi Hou}
\email{hou.shaoqi@whu.edu.cn}
\affiliation{School of Physics and Technology, Wuhan University, Wuhan, Hubei 430072, China}

\date{\today}

\begin{abstract}
  In this work, it is shown that based on the linear analysis, as long as a theory of gravity is diffeomorphism invariant and possesses the tensor degrees of freedom propagating at a constant, isotropic speed without dispersion, its asymptotic symmetry group of an isolated system contains the (extended/generalized) Bondi-Metzner-Sachs group.
  The tensor gravitational wave induces the displacement, spin and center-of-mass memory effects.
  They depend on the asymptotic shear tensor.
  The displacement memory effect is the vacuum transition and parameterized by a supertranslation transformation.
  All of these hold even when the Lorentz symmetry is broken by a special timelike direction.
\end{abstract}

\maketitle


\section{Introduction}

The gravitational memory effect has been studied in several theories of gravity together with the asymptotic symmetry group of the isolated system, including general relativity (GR) \cite{Zeldovich:1974gvh,Kovacs:1978eu,Braginsky:1986ia,1987Natur.327..123B,Christodoulou1991,Wiseman:1991ss,Blanchet:1992br,Thorne:1992sdb,Strominger:2014pwa,Flanagan:2015pxa,Pasterski:2015tva,Strominger:2018inf,Nichols:2018qac}, Brans-Dicke theory (BD)\cite{Hou:2020tnd,Tahura:2020vsa,Seraj:2021qja,Hou:2020wbo,Hou:2020xme,Tahura:2021hbk}, dynamical Chern-Simons theory (dCS) \cite{Hou:2021oxe,Hou:2021bxz}, and Einstein-\ae{}ther theory (E\AE) \cite{Hou:2023pfz}. 
These theories all contain at least two tensorial degrees of freedom (dofs), which excite the displacement, spin and center-of-mass (CM) memory effects in the tensor sector.
The asymptotic symmetry group of an isolated system in these theories share a common subgroup, the celebrated Bondi-Metzner-Sachs (BMS) group \cite{Bondi:1962px,Sachs:1962wk,Sachs1962asgr}, and its enlarged versions \cite{Barnich:2009se,Barnich:2010eb,Campiglia:2014yka,Campiglia:2015yka,Campiglia:2017xkp}.
In this group, supertranslations and (super-)Lorentz transformations \cite{Compere:2018ylh} are the nontrivial residual diffeomorphisms that preserve the asymptotic behaviors of the dynamical fields \cite{Strominger:2018inf}.
Supertranslations generalize the usual spacetime translations, transforming the vacuum state nontrivially in the tensor sector \cite{Ashtekar:1981hw}.
The displacement memory effect is actually the vacuum transition in the tensor sector, given by the change in the asymptotic shear.
The magnitude of the displacement memory is determined by the flux-balance law associated with the supertranslation \cite{Strominger:2014pwa,Strominger:2018inf,Hou:2020wbo,Hou:2021bxz}.
Likewise, the spin and CM memories are constrained by the flux-balance laws of the super-spin and super-rotation transformations, the magnetic and electric parity components of the super-Lorentz generator, respectively \cite{Pasterski:2015tva,Nichols:2017rqr,Nichols:2018qac,Tahura:2020vsa,Hou:2020wbo,Hou:2021bxz}.
Although the flux-balance laws have not yet been established in Ref.~\cite{Hou:2023pfz} based on the linear analysis, one expects that similar conclusions also hold in E\AE.
Indeed, a recent work proved that for a dynamical metric theory, the displacement memory effect in the tensor sector is always sourced by the energy flux of all radiative dofs in this theory \cite{Heisenberg:2023prj}.

Noticing the commonalities of the asymptotic symmetry group and the gravitational memory effect shared by these theories of gravity, one may conjecture that any theory of gravity that is diffeomorphism invariant, and predicts the existence of tensorial dofs, propagating at a constant, isotropic speed without any dispersion effect, would also share these commonalities.
Clearly, this conjecture holds in GR, BD, dCS and E\AE{}.
In this work, we will prove this conjecture more generally.
Given the fact that there are great variations in the form of Lagrangian, and inspired by Refs.~\cite{Blanchet:1986dk,Blanchet:2020ngx,Hou:2023pfz}, we would like to work in the linearized regime.
This will greatly simplify the derivation, and yet, produce certain results that would still be correct in the full, nonlinear theory.
This is because the memory effect occurs at large distances from the source of gravity, where the gravitational interaction is weak, and the linear analysis can be performed. 
In addition, the asymptotic symmetry does not depend on the interactions.
The linear analysis also allows us to construct a generic quadratic action in the tensor sector, making the analysis model-independent.

This work is orgainzed in the following way. 
Section~\ref{sec-sym} reviews mainly how to define symmetries in a gravitating system.
Section~\ref{sec-act} constructs the quadratic action of a generic theory of gravity.
Then, the symmetry and the memory effect will be discussed in Section~\ref{sec-asy}.
The discussion is generalized in Section~\ref{sec-gen}. 
And finally, we discuss and summarize in Section~\ref{sec-con}.
In the following, the units is chosen such that $c=G=1$.

\section{What is a symmetry?}
\label{sec-sym}

Generally speaking, a symmetry is some transformation acting on physical fields that preserves some aspects of the physical system.
Note that one does not have to demand the invariance of everything of the physical system in order to define a symmetry, otherwise, the symmetry would be most likely trivial.
Familiar transformations include the coordinate transformations and the internal transformations.
The (general) coordinate transformations are usually performed on the gravitational systems. 
The internal transformations are usually for the physical systems in the absence of gravity, such as the phase rotation of the wavefunction in quantum mechanics \cite{Cohen-Tannoudji:101367}, and many gauge transformations, e.g., U(1), SU(2), and etc., in quantum field theory \cite{Weinberg:1995mt}.
Of course, one would also perform the Lorentz transformation on physical systems not influenced by gravity.
In this work, the focus will be on gravity, and the internal transformation will not be considered further.

In theories of gravity, it is often required that under the coordinate transformation, the action of a gravitating system shall be invariant.
This gives the symmetry at the level of the action, which means that the physical fields, such as the metric in particular, may not be preserved under the coordinate transformation.
In some special physical systems, the physical fileds may be preserved under some coordinate transformations. 
Then, these coordinate transformations are called the (exact) symmetries of the physical system.
Since the metric is invariant, these symmetries are generated by some Killing vector fields \cite{Wald:1984rg}.
The Killing equations are nontrivially satisfied, so in the most gravitating systems, there are no Killing vectors, and only the symmetry of the action can be demanded. 

The absence of the exact symmetries of the most physical systems leads to the inablity to define conserved quantities (e.g., the energy and the momentum).
Therefore, certain approximate symmetries were pursued for suitable physical systems.
For example, in the weak field regime, the Poincar\'e symmetry of the background Minkowski metric is often invoked to define the energy, the momentum and the angular momentum of the gravitational wave emitted by a compact binary system \cite{Poisson2014}, but it is not a symmetry of the full metric. 
Another kind of appropriate symmetries is the asymptotic symmetry of an isolated system, which approximately preserves the metric near the boundary of the spacetime \cite{Bondi:1962px,Sachs:1962zza,Ashtekar:1978zz,Beig:1982cmp,Beig:1984int,Wald:1984rg}.
If one is interested in the geometry near the null infinity, the asymptotic symmetry is the BMS symmetry \cite{Bondi:1962px,Sachs:1962zza}; 
if the spatial infinity is considered, the asymptotic symmetry is the SPI symmetry \cite{Ashtekar:1978zz,Beig:1982cmp,Beig:1984int}.
Within the conformal completion method \cite{Penrose:1962ij}, the BMS symmetry can also be taken to be the transformation that keeps $\Gamma^{\mu\nu}{}_{\rho\sigma}=\hat g_{\rho\sigma}\hat n^\mu\hat n^\nu$, evaluated at the null infinity, invariant, where $\hat g_{\mu\nu}=\Omega^2g_{\mu\nu}$ is the conformal metric with $\Omega$ the conformal factor, and $\hat n^\mu=\hat g^{\mu\nu}\nabla_\nu\Omega$ \cite{Geroch1977}.
The SPI symmetry is actually the transformation that leaves unchanged the conformal metric $\hat g_{\mu\nu}$ at the spatial infinity, and the equivalence classes of spatial curves sharing the same 4-acceleration at the spatial infinity \cite{Ashtekar:1978zz,Prabhu:2019daz}.
The examples of the BMS and the SPI symmetries suggest that one may require some aspects of the unphysical metric together with other relevant fields to be invariant in order to define the approximate symmetry.
More recently, the generalized BMS symmetry was proposed, which does not even preserve the leading behavior of the physical $g_{\mu\nu}$, but keeps $\hat\epsilon_{\mu\nu\rho}\hat n^\alpha\hat n^\beta\hat n^\sigma$ invarint with $\hat\epsilon_{\mu\nu\rho}$ the volumn element of the null infinity \cite{Campiglia:2014yka,Campiglia:2015yka,Flanagan:2019vbl}.
Finally, the Weyl-BMS symmetry is to preserve the generators of the null infinity \cite{Freidel:2021fxf,Chandrasekaran:2021vyu}.
So the metric does not even play any explicit role in the definition of the Weyl-BMS symmetry.

From these examples, the symmetry of a physical system can be defined to preserve some aspects of the physical system in question.
The invairance of the physical metric $g_{\mu\nu}$ is not always demanded, either globally or asymptotically.
Therefore, in the following, after analyzing the dynamical equations of the metric perturbation, we propose the symmetry that is for the unphysical metric as illuated below. 
With the very symmetry defined, one can understand, in a unified manner, the tensorial memory effects of the modified theories of gravity satisfying the conditions of the conjecture metioned in Introduction.

\section{Quadratic action for a generic theory of gravity} 
\label{sec-act}

Here, a quadratic action will be constructed for a generic theory of gravity in the linear regime around the flat spacetime background, which is given by
\begin{equation}
  \label{eq-lag}
  \mathcal L=\mathcal L(h_{\mu\nu},\pd_{\rho}h_{\mu\nu};\psi,\pd_{\rho}\psi),
\end{equation}
where $h_{\mu\nu}=g_{\mu\nu}-\eta_{\mu\nu}$ is the metric perturbation. 
$\psi$ collectively represents other gravitational field, and its indices have been suppressed.
It can be a scalar field $\psi^I$ with $I$ indicating its species, or a vector field $\psi^I_\mu$.
$\psi$ could also be a rank-2 tensor, $\psi_{\mu\nu}^I$.
If $\psi^I_{\mu\nu}=\psi^I_{\nu\mu}$,  its kinetic interaction with $h_{\mu\nu}$ is required to be diagonalizable. 
The physical graviton is the linear combination of $h_{\mu\nu}$ and $\psi^I_{\mu\nu}$.
For simplicity, we assume that the kinetic interaction has already been diagonalized, and $h_{\mu\nu}$ is the physical graviton field.
Then, $\psi^I_{\mu\nu}$ interacts with $h_{\mu\nu}$ via terms like $\psi^I_{\mu\nu}h^{\mu\nu}$ or $\eta^{\mu\nu}\psi^I_{\mu\nu}\eta^{\rho\sigma}h_{\rho\sigma}$.
So $\psi^I_{\mu\nu}$ may play the rule of a second metric as in bimetric theories of gravity \cite{deRham:2014zqa,Schmidt-May:2015vnx}, and mass terms might be generated, leading to the dispersion effect.
Therefore, we forbid the interaction between $h_{\mu\nu}$ and $\psi^I_{\mu\nu}$. 
Effectively, we ignore the symmetric rank-2 tensor field $\psi^I_{\mu\nu}$. 
We do not consider the higher-rank tensors, as there are some difficulties in constructing a healthy higher-rank tensor theory with interactions \cite{Ponomarev:2022vjb}.
In summary, $\mathcal L$ contains $h_{\mu\nu}$, scalar fields $\psi^I$, vector fields $\psi^I_\mu$, and the antisymmetric rank-2 tensor fields $\psi^I_{[\mu\nu]}$, which generally couple with each other.
There is no constraint on the number of gravitational fields other than $h_{\mu\nu}$.
Being a quadratic action, $\mathcal L$ is quadratic in  $h_{\mu\nu}$ and $\psi$ and their derivatives.
The possibility of incorporating higher order derivatives will be considered later.

One demands $\mathcal L$ be diffeomorphism invariant, i.e., invariant under an infinitesimal coordinate transformation, $x^\mu\rightarrow x^\mu+\xi^\mu$, in the linear regime.
In this section, one should also assume the Lorentz invariance is respected, or if not, it is violated by the presence of a special timelike direction.
Under this assumption, one knows that the spatial rotational symmetry always exists.
This suggests to write $\mathcal L$ in the 3+1 formalism in a particular Lorentz frame with the time coordinate parallel to the special timelike direction.
Since $\mathcal L$ is required to be diffeomorphism invariant, it should be expressed in terms of gauge-invariant variables.
It is well-known that one can construct several gauge invariants from $h_{\mu\nu}$ \cite{Flanagan:2005yc} by first rewriting it in the following way,
\begin{gather*}
  h_{tt} = 2\phi,\\
  h_{tj} = \beta_j+\partial_j\zeta,\\
  h_{jk} = h_{jk}^\mathrm{TT}+\frac{1}{3}H\delta_{jk}+\partial_{(j}\varepsilon_{k)}+\left(\partial_j\partial_k-\frac{\delta_{jk}}{3}\nabla^2\right)\rho,
\end{gather*}
with $\pd^kh_{jk}^\mathrm{TT}=0$, and $\delta^{jk}h_{jk}^\mathrm{TT}=\pd^j\beta_j=\pd^j\varepsilon_j=0$, and then forming the combinations 
\begin{gather*}
  \Phi=\pd_t \zeta-\phi-\frac{1}{2}\pd_t^2\rho,\quad \Theta=\frac{1}{3}(H-\nabla^2\rho),\\
  \Xi_j=\beta_j-\frac{1}{2}\pd_t\varepsilon_j,\\
  h_{jk}^\text{TT},
\end{gather*}
which are gauge-invariant variables.
There might be more gauge invariants built from $\psi$ to be discussed below.

In order to define gauge invariants based on $\psi$, one should first consider the vacuum expectation values (vevs) of $\psi$.
For a scalar field, the vev $\langle\psi^I\rangle$ shall still be a constant.
If there are several nontrivial vevs for vector fields, they shall be parallel to each other, i.e., $\langle\psi_\mu^I\rangle\propto\delta_\mu^0$.
Finally, if the antisymmetric tensor field $\psi_{\mu\nu}^I$ has a nonvanishing background value, one can check that its eigenvectors are all null.
So if $\psi_{\mu\nu}^I$ is present in $\mathcal L$, its vev shall vanish.
Therefore, in order to break the Lorentz invariance, there must exist at least one vector field $\psi_\mu^I\equiv v_\mu^I$, and its vev $\langle v_\mu^I\rangle=v^I\delta^0_\mu$.
Thus, a scalar field $\psi^I$ is gauge invariant.
For the antisymmetric tensor field $\psi_{\mu\nu}^I$, define vectors 
\begin{equation*}
 \chi_j^I=\psi^I_{0j},\quad \lambda_j^I=\epsilon^{jkl}\psi_{kl}^I, 
\end{equation*}
both of which are gauge invariant.
If $\psi$ represents a vector field whose vev is trivial, it is gauge invariant, too. 
The situation is complicated when $\psi$ stands for $v_\mu^I$, whose vev does not vanish.
One should decompose $v_j^I$ such that 
\begin{equation*}
v_j^I=\mu_j^I+\pd_j\nu^I, \quad\pd^j\mu_j^I=0.
\end{equation*}
Then, the gauge invariants are $\mu_j^I$ and $\Pi^I=v_0^I-\pd_t \nu^I$.

In summary, the gauge invariants include several scalars (e.g., $\Phi$, $\Theta$, $\Pi^I$, and $\psi^I$) and vectors (e.g., $\Xi_j$, $\mu_j^I$, $\chi_j^I$, $\lambda_j^I$, and $\psi^I_j$ with a vanishing vev).
There is only one rank-2 tensor, $h_{jk}^\text{TT}$.
This has a profound consequence, that is, $h_{jk}^\text{TT}$ decouples from the remaining gauge invariants in $\mathcal L$.
This is because the rotational symmetry implies that what shall appear in $\mathcal L$ are terms whose time indices shall be contracted with time indices, and the space indices shall be contracted with space indices.
For example, terms like $\pd_t  h_{jk}^\mathrm{TT}\pd_t h^{jk}_\mathrm{TT}$, $\pd_lh_{jk}^\mathrm{TT}\pd^lh^{jk}_\mathrm{TT}$,  $\pd_{j}\Xi_k\pd^j\Xi^k$, $\pd_j\Phi\pd^j\Phi$, and $\pd_j\Theta\pd^j\Phi$ are allowed to exist in $\mathcal L$.
One could also write down the similar contractions of the remaining gauge invariants.
Note that there should be no terms such as $\pd_lh_{jk}^\mathrm{TT}\pd^jh^{kl}_\mathrm{TT}$, or $\pd_j\Xi_k\pd^k\Xi_j$, because suitable integration by parts removes them.
There should be no contractions of different tensor types, either.
For example, no terms like $\pd_lh_{jk}^\mathrm{TT}\pd^l\Xi^j$, $\pd_j\Xi_k\pd^j\Phi$ or $\pd_lh_{jk}^\mathrm{TT}\pd^l\Phi$ shall exist in $\mathcal L$.
These terms shall be contracted with constant coefficients carrying space indices in order to produce scalar quantities. 
For the first two terms, the coefficient has the form of a 3-vector, $c^k$, which selects a preferred spatial direction.
This case shall be excluded.
For the last term, $\pd_lh_{jk}^\mathrm{TT}\pd^l\Phi$, a coefficient of the form $c'^{jk}$ shall be contracted with it.
In order to maintain the rotational symmetry, $c'^{jk}\propto\delta^{jk}$, but $h_{jk}^{\text{TT}}$ is traceless.

The above arguments suggest that $\mathcal L$ has the following part for $h_{jk}^\mathrm{TT}$,
\begin{equation}
  \label{eq-tt-lag}
  \mathcal L_h\propto \frac{1}{s_2^2}\pd_t h_{jk}^\mathrm{TT}\pd_t h^{jk}_\mathrm{TT}-\pd_lh_{jk}^\mathrm{TT}\pd^lh^{jk}_\mathrm{TT}.
\end{equation}
Here, $s_2$ is a constant.
If $s_2=1$, the tensor sector respects the Lorentz invariance, otherwise, the Lorentz symmetry is broken.
The remaining part of $\mathcal L$ can take a more complicated form, as long as it respects the diffeomorphism invariance, is Lorentz invariant or violating by a special timelike direction, and is healthy (e.g., without ghosts).
Since it is decoupled from $\mathcal L_h$, it is irrelevant for the following discussion.
The Lagrangian $\mathcal L_h$ appears in many special theories of gravity, including GR \cite{Wald:1984rg}, BD \cite{Brans:1961sx,Eardley:1974nw}, $f(R)$ theory \cite{Liang:2017ahj}, Horndeski theory \cite{Horndeski:1974wa,Hou:2017bqj}, Einstein-Gauss-Bonnet theory \cite{Kobayashi:2011nu}, and E\AE{} \cite{Jacobson:2000xp,Jacobson:2004ts,Gong:2018cgj},  etc.

\section{Asymptotic analysis and the memory effect} 
\label{sec-asy}

The equation of motion derived from the action~\eqref{eq-tt-lag} is
\begin{equation}
  \label{eq-eom}
  -\frac{1}{s_2^2}\frac{\pd^2}{\pd t^2} h_{jk}^\mathrm{TT}+\nabla^2h_{jk}^\mathrm{TT}=0.
\end{equation}
So $s_2$ is the speed of the tensor gravitational wave (GW).
It is simpler to perform the asymptotic analysis if $\mathcal L$ respects the Lorentz invariance, following the methods in Refs.~\cite{Blanchet:1985sp,Blanchet:2020ngx}.
If $\mathcal L$ violates the Lorentz symmetry, then $s_2\ne1$, and the causal structure of $h_{jk}^\text{TT}$ differs from the one determined by $\eta_{\mu\nu}$.
Inspired by Ref.~\cite{Hou:2023pfz}, one may perform the following disformal transformation to get the unphysical metric $\eta'_{\mu\nu}$
\begin{equation}
  \label{eq-disf}
  \begin{split}
  \eta_{\mu\nu}\rightarrow\eta'_{\mu\nu}&=\eta_{\mu\nu}+(1-s_2^2)\delta^0_\mu\delta^0_\nu\\
  &=\mathrm{diag}\{-s_2^2,1,1,1\},
  \end{split}
\end{equation}
then, Eq.~\eqref{eq-eom} becomes  
$$\eta'^{\mu\nu}\pd_\mu\pd_\nu h_{jk}^\mathrm{TT}=0,$$ 
where $\eta'^{\mu\nu}$ is the inverse of $\eta'_{\mu\nu}$.
Therefore, the dynamics and the causal structure of $h_{jk}^\text{TT}$ are determined by $\eta'_{\mu\nu}$, instead of $\eta_{\mu\nu}$.
Once one carries out a coordinate transformation $x^\mu=(t,\vec x)\rightarrow x^{\bar\mu}=(\bar t=s_2t,\vec x)$, the above equation becomes,
\begin{equation}
  \label{eq-eom-2}
  -\frac{\pd^2}{\pd\bar t^2}h_{jk}^\text{TT}+\nabla^2h_{jk}^\text{TT}=0,
\end{equation}
a familiar form as in GR.
Formally, $\eta'_{\mu\nu}$ plays the role of a metric, and one may call it unphysical.
In the unphysical spacetime associated with $\eta'_{\mu\nu}$, $h_{jk}^\text{TT}$ is still transverse-traceless, and would travel along the null direction defined by $\eta'_{\mu\nu}$.
In addition, Lorentz invariance is recovered with respect to $\eta'_{\mu\nu}$.
Similarly, $\mathcal L_h$ can be rewritten as 
\begin{equation}
  \label{eq-tt-lag-1}
  \mathcal L_h\propto \pd_{\bar t}h_{jk}^\text{TT}\pd_{\bar t}h^{jk}_\text{TT}-\pd_lh_{jk}^\text{TT}\pd^lh^{jk}_\text{TT}.
\end{equation}
Although this is expressed in the 3+1 form, it is diffeomorphism invariant, as $h_{jk}^\text{TT}$ is a gauge-invariant variable, under the infinitesimal coordinate transformation given by $\xi^\bmu=(\xi^{\bar t},\xi^j)=(s_2\xi^t,\xi^j)$.
So it is possible to rewrite Eq.~\eqref{eq-tt-lag-1} in a manifestly covariant form.
Define a tensor $\tilde h_{\bmu\bnu}^\text{TT}$ such that $\tilde h_{\bar t\bmu}^\text{TT}=0$, and $\tilde h_{jk}^\text{TT}=h_{jk}^\text{TT}$ in the current Lorentz frame.
Then, Eq.~\eqref{eq-tt-lag-1} becomes 
\begin{equation}
  \label{eq-tt-lag-2}
\mathcal L_h\propto \pd_{\brho}\tilde h_{\bmu\bnu}^\text{TT}\pd^{\brho}\tilde h^{\bmu\bnu}_\text{TT},
\end{equation}
which does not retain this form if a gauge transformation is performed. 
Here, $\eta'^{\bmu\bnu}$ is used to raise the indices.
One has to rewrite the right-hand side of Eq.~\eqref{eq-tt-lag-2} further, that is, \cite{Carroll:2004st}
\begin{equation}
  \label{eq-tt-lag-3}
  \mathcal L_h\propto \pd_\brho\tilde h_{\bmu\bnu}\pd^\brho\tilde h^{\bmu\bnu}-2\pd_\brho\tilde h_{\bmu\bnu}\pd^\bmu \tilde h^{\brho\bnu}+2\pd_\bmu \tilde h^{\bmu\bnu}\pd_\bnu\tilde h-\pd^\bmu\tilde h\pd_\bmu \tilde h,
\end{equation}
where $\tilde h_{\bmu\bnu}=\tilde h_{\bmu\bnu}^\text{TT}-2\pd_{(\bmu}\xi_{\bnu)}$, and $\tilde h=\eta'^{\bmu\bnu}\tilde h_{\bmu\bnu}$.
To get this expression, one simply writes down the linear combination of all possible contractions of derivatives of $\tilde h_{\bmu\bnu}$, requires the diffeomorphism invariance and makes sure that the resultant Lagrangian reduces to Eq.~\eqref{eq-tt-lag-2} in the transverse-traceless gauge \cite{Schwartz:2013pla}.
Generally speaking, $\tilde h_{\bmu\bnu}$ is no longer transverse-traceless in order to make the diffeomorphism invariance manifest.
Equation~\eqref{eq-tt-lag-3} is nothing but the quadratic Lagrangian of the Einstein-Hilbert action \cite{Carroll:2004st}.

So now, one effectively recovers GR's quadratic action, embedded in a generic theory of gravity that satisfies the conditions in the conjecture.
Ho\v{r}ava gravity \cite{Horava:2009uw}, and the alike, cannot contain a quadratic Lagrangian like Eq.~\eqref{eq-tt-lag-3} in the tensor sector, as they are not diffeomorphism invariant.
Then, one could directly apply the treatments in Refs.~\cite{Blanchet:1985sp,Blanchet:2020ngx} to analyze the asymptotic behaviors of $\tilde h_{\bmu\bnu}$.
That is, one first fixes the gauge, e.g., imposing the harmonic gauge $\pd^\bnu \bar h_{\bmu\bnu}=0$ with $\bar h_{\bmu\bnu}=\tilde h_{\bmu\bnu}-\eta'_{\bmu\bnu}\tilde h/2$.
Then, $\bar h_{\bmu\bnu}$ satisfies $\eta'^{\brho\bsigma}\pd_\brho\pd_\bsigma \bar h_{\bmu\bnu}=0$.
A generic solution to this equation in the Lorentz frame $\{x^\bmu\}$ can be written in terms of $M_L$ and $S_L$, the mass and current multipole moments, respectively \cite{Blanchet:2020ngx}.
Then, a suitable coordinate transformation, $x^\bmu\rightarrow x^{\tmu}=(\tdu,\tdr,\tdth^a)$ with $\tdth^a=\tdth,\hat\varphi$, can be performed such that the perturbed metric $g'_{\tmu\tnu}=\eta'_{\tmu\tnu}+\tilde h_{\tmu\tnu}$ satisfies $g'_{\tdu\tdr}=-1$ and $g'_{\tdr\tdr}=g'_{\tdu \ta}=0$ (Newman-Unti gauge \cite{Newman:1962cia}).
Thus, at the leading order in the field perturbation, 
\begin{gather}
	\tdu=s_2t-r+\xi^{\tdu}+\cdots,\\
\tdr=r+\xi^\tdr+\cdots,\\
\tdth^a=\theta^a+\xi^{\ta}+\cdots,
\end{gather}
with $r=|\vec x|$.
Here, 
\begin{equation}
	\xi^{\tdu}=f,\quad \xi^{\tdr}=-\tdr f^{[1]}+Q,\quad\xi^\ta=Y^\ta-\frac{1}{\tdr}\tdcd^\ta f,
\end{equation}
in which $f$, $Q$, and $Y^\ta$ are arbitrary functions of $\tdu$ and $\tdth^a$, and $\tdcd_\ta$ is the covariant derivative of the leading order part of $\tdr^{-2}g'_{\hat a\hat b}\equiv\tdgamma_{\ta\tb}+\cdots$.
Requiring that 
\begin{equation}
	g'_{\tdu\tdu}=\order{\tdr^0},\quad g'_{\tdu\ta}=\order{\tdr^0},\quad\det(g'_{\ta\tb})=\tdr^4\sin^2\tdth+\order{\tdr^2},
\end{equation}
one knows that 
\begin{equation}
	f=T+\frac{\tdu}{2}\tdcd_\ta Y^\ta,\quad Q=\frac{1}{2}\tdcd^2 f,
\end{equation}
with $\tdcd^2=\tdcd_\ta\tdcd^\ta$, and both of $T$ and $Y^\ta$ are independent of $\tdu$.
So $T$ is the supertranslation generator, and $Y^\ta$ generates the diffeomorphisms on the topological 2-sphere.
Then, one knows that $\xi^\tmu$ is in the generalized BMS algebra \cite{Campiglia:2017xkp,Campiglia:2018see,Campiglia:2020qvc}.
If one further requires that $\gamma_{\ta\tb}$ be the round metric on the unit 2-sphere, $Y^\ta$ is now the conformal Killing vector field of $\gamma_{\ta\tb}$, so the asymptotic symmetry group is the standard BMS group.
Of course, if one allows $Y^\ta$ to have  a finite number of singularities, one gets the extended BMS group \cite{Barnich:2009se,Barnich:2010eb,Barnich:2011mi,Flanagan:2015pxa}.
Therefore, all of these statements on the asymptotic symmetry are consistent with the conjecture.

The subleading order of $g'_{\ta\tb}$ is determined by the asymptotic shear tensor $c_{\ta\tb}$, and a nonvanishing news $N_{\ta\tb}\equiv \pd_{\tdu}c_{\ta\tb}$ indicates the presence of the tensor GW at $\tdr\rightarrow\infty$.
Under a supertranslation $T$, 
\begin{equation}
c_{\ta\tb}\rightarrow c_{\ta\tb}-2\tdcd_{\langle\ta}\tdcd_{\tb\rangle}T,
\end{equation}
where the square brackets mean to take the traceless part with respect to $\gamma_{\ta\tb}$.
Now, consider the memory effect excited by the tensor modes.
Let $D^j$ be the separation between adjacent test particles and $\tau$ be the affine parameter.
The geodesic deviation equation 
\begin{equation}
\frac{\ud^2D^j}{\ud\tau^2}=-R_{tjtk}D^k
\end{equation}
can be integrated to determine the memory effects. 
For that end, it shall be rewritten in the new coordinate system $x^{\hat\mu}$.
Since we are interested in the tensor memory effects, $R_{tjtk}=-\pd_t^2h_{jk}^\text{TT}/2$, and the angular components are 
\begin{equation}
\frac{\ud^2D^\ta}{\ud\tau^2}=-\frac{s_2^2}{\tdr^2}\tdgamma^{\ta\tb}R_{\tdu\tb\tdu\tc}D^\tc.
\end{equation}
Following Ref.~\cite{Hou:2023pfz}, it is easy to show that at the leading order in $\tdr$, $R_{\tdu\ta\tdu\tb}\propto\pd_{\tdu}^2 c_{\ta\tb}$.
So one defines the displacement memory effect to be 
\begin{equation}
\Delta c_{\ta\tb}=c_{\ta\tb}(\tdu_f)-c_{\ta\tb}(\tdu_i),
\end{equation}
where the GW exists between $\tdu_i$ and $\tdu_f$.
When the GW is absent, the system is said to be in the vacuum state.
In fact, one can show that $\Delta c_{\ta\tb}=-2\tdcd_{\langle\ta}\tdcd_{\tb\rangle}T$ for some suitable $T$.
So the displacement memory effect is the vacuum transition, parameterized by a supertranslation as stated in the conjecture.
Define $c_{\ta\tb}=\tdcd_{\langle\ta}\tdcd_{\tb\rangle}\mathcal E+\epsilon_{\tc\langle\ta}\tdcd_{\tb\rangle}\tdcd^\tc\mathcal M$.
The spin memory is quantified by $\int_{\tdu_i}^{\tdu_f}\mathcal M\ud\tdu$, while the CM memory by $\int_{\tdu_i}^{\tdu_f}\tdu\pd_{\tdu}\mathcal E_o\ud\tdu$, where $\mathcal E_o$ is the part of $\mathcal E$ given by $\Delta m=\tdcd^2(\tdcd^2+2)\Delta\mathcal E_o/8$ and $m$ is the half of the coefficient of the subleading part of $g'_{\tdu\tdu}$, i.e., the Bondi mass aspect.
Therefore, all three memory effects in the tensor sector depend on the evolution of $c_{\ta\tb}$ in agreement with the conjecture.

As in Ref.~\cite{Hou:2023pfz}, the linearized analysis cannot produce the flux-balance laws, so it is not possible to associate various flux-balance laws with these memory effects. 
However, if one considers the nonlinear corrections, one expects that the displacement memory effect is constrained by the flux-balance law corresponding to the supertranslation, as verified by \cite{Heisenberg:2023prj} for the so-called dynamic metric theories, which are special cases considered in this work.
Moreover, it should also be correct that the spin memory effect is related to the flux-balance law of the super-rotations, and the CM memory can be calculated using the flux-balance law of the super-boosts.
These will be considered in the future.

\section{Generalization}  
\label{sec-gen}

The Lorentz violation would also be caused by the absence of the rotational symmetry.
One way to achieve this is to introduce some special spacelike vector fields, so the isotropy of the space is broken, as in the bumblebee gravity \cite{Kostelecky:2003fs}.
Or, there could be a nonvanishing vev of some tensor field, whose eigenvectors define preferred timelike directions or spacelike directions.
The standard-model extension (SME) provides a great amount of such tensor fields \cite{Kostelecky:2003fs,Kostelecky:2016kfm}.
Finally, if there are two nontrivial timelike vector fields, not parallel to each other, the rotational invariance is also violated.
However, such a theory  may predict that the GW solution would travel at a speed dependent on the propagation direction, and the dispersion generally takes place, e.g., in SME \cite{Kostelecky:2016kfm}.
In some special cases, the dispersion effect does not occur, but additional modes (e.g. $\Phi$, $\Theta$, or $\Xi_j$) could be excited by $h_{jk}^\text{TT}$ \cite{Hou:2024xbv}.
This is because whenever the rotational symmetry is broken, $h_{jk}^\text{TT}$ is allowed to couple with the scalar and vector gauge invariants, e.g., $c^k_1\pd_lh_{jk}^\mathrm{TT}\pd^l\Xi^j$, $c_2^k\pd_j\Xi_k\pd^j\Phi$ or $c_3^{jk}\pd_lh_{jk}^\mathrm{TT}\pd^l\Phi$ with $c_1^k,c_2^k,c_3^{jk}(\ne\delta^{jk}c_3^{il}\delta_{il}/3)$ constant.
A nonvanishing $\Phi$, $\Theta$ or $\Xi_j$ would alter the asymptotic behaviors of $g_{\mu\nu}$ a lot, and thus the asymptotic symmetry and memory effects.

If it happens that $h_{jk}^\text{TT}$ does not couple with the other fields, and satisfies $\mathfrak K^{\mu\nu}\pd_\mu\pd_\nu h_{jk}^\text{TT}=0$ 
with $\mathfrak K^{\mu\nu}(\ne\eta^{\mu\nu}\mathfrak K_{\rho\sigma}\eta^{\rho\sigma}/4)$ constant, then the causal structure of $h_{jk}^\text{TT}$ is determined by a more general tensor $\mathfrak K^{\mu\nu}$ instead of $\eta_{\mu\nu}$, or $\eta'_{\mu\nu}$.  
So in this case, the inverse $\mathfrak K_{\mu\nu}$ of $\mathfrak K^{\mu\nu}$ is viewed to be the unphysical metric, similar to $\eta'_{\mu\nu}$ introduced previously.
In a new coordinate system $x^{\hat\mu}=\Lambda^{\hat\mu}{}_\nu x^\nu$, it is possible that $\mathfrak K^{\hat\mu\hat\nu}=\text{diag}\{-1,1,1,1\}$ for convenience, and one has  
\begin{equation*}
  \mathfrak K^{\trho\tsigma}\pd_\trho\pd_\tsigma h_{\tmu\tnu}^\text{TT}=-\frac{\pd^2}{\pd\hat t^2}h_{\tmu\tnu}^\text{TT}+\frac{\pd^2}{\pd x^{\hat j}\pd x^{\hat j}}h_{\tmu\tnu}^\text{TT}=0,
\end{equation*}
where $h_{\tmu\tnu}^\text{TT}=\tilde\Lambda^\rho{}_\tmu\tilde\Lambda^\sigma{}_\tnu h_{\rho\sigma}^\text{TT}$ with $\tilde\Lambda^\rho{}_\tmu\Lambda^\tnu{}_\rho=\delta^\tnu_\tmu$.
Unfortunately, in the unphysical spacetime, $h^\text{TT}_{\tmu\tnu}$ is no longer transverse-traceless, i.e., $\mathfrak K^{\trho\tnu}\pd_{\trho}h^\text{TT}_{\tmu\tnu}\ne0$ and $\mathfrak K^{\tmu\tnu}h_{\tmu\tnu}^\text{TT}\ne0$, in general.
However, one could uniquely construct a transverse-traceless tensor $\mathfrak h_{\tmu\tnu}^\text{TT}$ from $h_{\tmu\tnu}^\text{TT}$, following the standard procedure \cite{Flanagan:2005yc}.
Note that $\mathfrak h_{\tmu\tnu}^\text{TT}$ still contains two tensorial dofs as $h_{\tmu\tnu}^\text{TT}$, which satisfies
\begin{equation}
  \label{eq-goh}
\mathfrak K^{\trho\tsigma}\pd_\trho\pd_\tsigma \mathfrak h_{\tmu\tnu}^\text{TT}=0,
\end{equation}
like Eq.~\eqref{eq-eom-2}.
This suggests that one could use $\mathfrak h_{\tmu\tnu}^\text{TT}$ to define the asymptotic symmetry group and the  memory effect, similar to the discussion in the previous section.
Therefore, in this very special case with the rotational symmetry broken, the asymptotic symmetry and the memory effect in the tensor sector are also the same as in GR.

\section{Discussion and conclusion} 
\label{sec-con}

The Lagrangian \eqref{eq-lag} considered previously contains no higher order derivatives of $h_{\mu\nu}$ and $\psi$.
Many modified theories of gravity have higher order derivatives in the nonlinear Lagrangian, such as Horndeski theory, higher derivative gravity \cite{Stelle:1976gc}, and E\AE{}.
Within the current framework, it is easy to incorporate the higher derivative terms in Eq.~\eqref{eq-lag}.
In fact, a similar argument shows that $h_{jk}^\text{TT}$ still does not interact with higher derivative terms of the remaining gauge invariants in the quadratic action, if the space is still rotationally invariant.
When there are higher order derivatives of $h_{jk}^\text{TT}$, they shall be highly suppressed by small coupling constants, so their effects on the dynamics of $h_{jk}^\text{TT}$ will be ignored.
Therefore, when there are higher order derivatives in the action, the Lagrangian for $h_{jk}^\text{TT}$ is still given by Eqs.~\eqref{eq-tt-lag} or \eqref{eq-tt-lag-3}, and the conjecture holds, too.

In summary, based on the linear analysis, if a theory of gravity is diffeomorphism invariant and possesses two tensorial dofs that move at a constant, isotropic speed without dispersion, the asymptotic symmetry group is the familiar BMS group or its enlarged versions in a suitably defined unphysical spacetime. 
The memory effects induced by the tensor modes include the displacement, spin and CM memories.
They are all functions of the asymptotic shear. 
The displacement memory is parameterized by a supertranslation, representing the vacuum transition. 
The linearized analysis is permitted as memory effects take place at the asymptotic regions, and enable a general discussion.
The linear analysis does not give the flux-balance laws to constrain the various memory effects, let alone the associated soft theorems \cite{Weinberg:1965s,Strominger:2018inf}. 
One should extend the current analysis to include nonlinear terms in order to determine the constraints.
This might be done following the ideas in Ref.~\cite{Heisenberg:2023prj}, for instance.
It is expected that the displacement, spin and CM memories are constrained by flux-balance laws associated with the supertranslation, super-rotation and super-boost transformations, respectively.

\bigskip

\textit{Acknowledgements.}
The author was grateful for the discussion with Pengming Zhang.
S. H. was supported by the National Natural Science Foundation of China under Grant No.~12205222.



\bibliographystyle{apsrev4-2}
\bibliography{tensormemorymdpi.bbl}

\end{document}